\documentstyle[ltwol]{article}

\def\Journal#1#2#3#4{{#1} {\bf #2}, #3 (#4)}

\def\PRL{\em Phys. Rev. Lett.}

\def\be{\begin{equation}}
\def\ee{\end{equation}}
\def\bea{\begin{eqnarray}}
\def\eea{\end{eqnarray}}
\begin{document}

\title{GRBS: STANDARD MODEL \& BEYOND}

\author{T. LU}

\address{Department of Astronomy, Nanjing University, Nanjing 210093, China\\E-mail: tlu@nju.edu.cn}   


\twocolumn[\maketitle\abstracts{ There have been great and rapid progresses 
in the field of $\gamma$-ray bursts since 
BeppoSAX and other telescopes discovered their afterglows in 1997.
In this talk, the main observational facts of $\gamma$-ray bursts and 
their afterglows, and the standard fireball shock model are reviewed briefly. 
And then, various post-standard effects, 
deviations from the standard model, are presented.}] 

\section{Observational Features}
$\gamma$-ray burst (or shortly, GRB) was discovered 
by R.W. Klebesadel et al.~\cite{kl} in 1967 and published in 1973. Although more 
than 2000 GRBs have been discovered, they 
are still among the most mysterious astronomical 
objects even at present time~\cite{pi,lu,rs}.
 
The GRB duration ($T$) is very short, usually only a 
few seconds or tens of seconds, or occasionally as long as a few tens of minutes
or as short as a few milli-seconds. Their time profiles are diverse, may be 
rather complicated. The time scales of variability 
($\delta T$) may be only milli-seconds or 
even sub-milli-seconds. The photon energy radiated in GRB is typically in the range of tens 
keV to a few MeV. However, high energy tail up to GeV or even higher than 10 GeV 
does exist. The spectra are definitely non-thermal and can usually be fitted 
by power law $N(E)dE \propto E^{-\alpha} dE$ (or break power law) with 
$\alpha \sim 1.8$ to $2.0$. The cosmological distances of GRBs were first inferred
statistically by the BATSE observation of their highly isotropic spatial 
distribution~\cite{mg}. The definite results of their cosmological origin come
from the observations of the red-shifts of their host galaxies after
the discovery of afterglows by BeppoSAX and various Satellites and Observatories since 1997.
Thus, GRBs are known to be the most energetic events ever known since the Big Bang.

Afterglows are the counterparts of GRBs at wave bands other than 
$\gamma$-rays. They are variable, typically decaying according to power 
laws: $F_{\nu} \propto t^{-\alpha}$ 
($\nu=$ X, optical, ......) with $\alpha = 1.1 - 1.6$ for X-ray, $\alpha 
= 1.1 - 2.1$ for optical band. Their spectra appear to be also of power law or 
broken power law. X-ray afterglows can last days or even weeks; 
optical afterglows and radio afterglows months or even one year. The light curves
of afterglows are much more smooth and simple than that of GRBs themselves.

\section{The Standard Fireball Shock Model}
First of all, the millisecond variabilities indicate that the GRBs should 
be compact stellar events only. In other words, the size of their initial 
burst region should be less than 300 km and the mass of a GRB source should 
be less than 100 M$_\odot$. As GRBs are at cosmological distances, their 
radiation (if isotropic) should be about $10^{51}$ to $10^{54}$ ergs.
So large an energy contained in so small a volume, it must be a fireball. 
The free path of a photon in the fireball can be shown to be less than its 
radius by a factor of about $10^{14}$, it is highly
optically thick. Such a fireball must be a black body. However, the observed 
radiations from GRBs are non-thermal, they must not come directly from the 
fireball. So large an optical depth also means that its pressure must be very 
high, and the fireball will expand relativistically to be a shell. If its 
expansion reaches ultra-relativistic speed with Lorentz factor of more than 100, 
the shell can then begin to be optically thin, and the non-thermal radiation 
of GRB can be observed.~\cite{pi}

Now, people have obtained a simple standard picture of GRBs. The inner engine at the center 
of the GRB source can send out several shells impulsively. As shell expansion reaches 
ultra-relativistic speed, the late fast shell can catch up and collide with the 
early slow shell and produce shocks (known as internal
shocks), and all shells will finally sweep the interstellar medium and produce shocks (known
as external shocks). The electrons accelerated by shocks will emit synchrotron radiations. The
internal shocks appear at about $10^{13}$ cm from the center and give out GRBs, and the 
external shocks at about $10^{16}$ cm and give out afterglows.
 
\section{The Post-standard Effects}

The standard model described above is based on the following simple assumptions: 
(1) the fireball expanding relativistically and isotropically; (2) impulsive 
injection of energy from inner engine to the fireball(s); (3) synchrotron radiation 
as the main radiation mechanism; (4) uniform environment with typical particle number density 
of $n=1$ cm$^{-3}$. This model is rather successful in that its physical 
picture is clear, results obtained are simple, and
observations on GRB afterglows support it at least qualitatively but generally. However, 
various quantitative deviations have been found. Thus, the simplifications made in the 
standard model should be improved. These deviations may reveal important new information. 
In the following, we will present some work of our group on the effects due to these 
deviations (the post-standard effects).

The fireball expansion is initially ultra-relativistic and highly 
radiative, this phase has been well described by some simple scaling 
laws.~\cite{ms,vi,wa,wi} The key equation~\cite{bl,ch} here is
\begin{equation}
\frac{{\rm d}\gamma}{{\rm d}m} = -\frac{\gamma^{2}-1}{M},
\end{equation}
$m$ denotes the rest mass of the swept-up medium, $\gamma$ the bulk Lorentz 
factor, and $M$ the total mass in the co-moving frame including internal energy $U$. 
This equation was originally derived under ultra-relativistic condition. The 
widely accepted results derived under this equation are correct for 
ultra-relativistic expansion. Accidentally, these results are also suitable for the 
non-relativistic and radiative case. However, for the important non-relativistic and 
adiabatic case, they will lead to wrong result ``$v \propto R^{-3}$'' ($v$ and $R$ is the 
velocity and radius of the fireball), perfectly different from the famous 
Sedov result of ``$v \propto R^{-3/2}$'', 
as first pointed out by Huang, Dai and Lu~\cite{hu,ha}. It is important to note that 
a fireball will usually become non-relativistic and adiabatic only several days after 
the burst~\cite{hn,hg,we,da}, while the 
afterglows can remain observable for several months or even about one year, so 
any useful model
should be able to account for both relativistic and non-relativistic, and both 
radiative and adiabatic phases. In order for solving this problem, Huang, Dai \& Lu~\cite{hu,ha}
further pointed out that the above equation (1) should be replaced by 
\begin{equation}
\frac{{\rm d}\gamma}{{\rm d}m} = -\frac{\gamma^{2}-1}{M_{ej} + \epsilon m 
+ 2(1-\epsilon)\gamma m},
\end{equation}
here $M_{ej}$ is the mass ejected from GRB central engine, $\epsilon$ 
(assumed to be approximately a constant) is the 
radiated fraction of the shock generated thermal energy in the co-moving 
frame. This equation leads to correct results for all cases 
including the Sedov limit for the non-relativistic and adiabatic phase.~\cite{hu,va}  

In the early days after the discovery of afterglows, Dai and Lu~\cite{zz} studied the 
possible non-uniformity of the surrounding medium. They used the general form of 
$n \propto R^{-k}$ to describe the non-uniform environment number density. By fitting 
the X-ray afterglow of GRB970616, they found $k=2$ which is just the form of a wind 
environment. This indicates that the surrounding medium of GRB970616 was 
just a stellar wind. After the detailed studies by Chevalier 
and Li~\cite{ce,cv}, the stellar wind model for the environment of
GRBs has now become widely interested. As the properties of GRBs' environment contain 
important information related with their pregenitors, this stellar wind model 
provides strong support to the view of massive star origin of GRBs.

Another environment effect is due to the deviation from the standard number density of 
$n=1$ cm$^{-3}$. Some afterglows of GRBs show that their lightcurves obey a broken power law. 
For example, according to Fruchter et al.~\cite{fr}, the optical lightcurve of GRB990123 
shows a break after about two days, its slope being steepened from -1.09 to -1.8. Dai and 
Lu~\cite{di} pointed out that a shock undergoing the transition from a relativistic phase to 
a non-relativistic phase may show such a break in the light curve. If there are 
dense media and/or clouds in the way, this break may happen earlier to fit the observed 
steepening. Recently, Wang, Dai \& Lu~\cite{wn} proved that the dense environment model can also 
explain well the radio afterglow of GRB 980519~\cite{fa}.

Lightcurves of some optical afterglows even show the 
down-up-down variation such as GRB970228 and 970508. These features can be 
explained by additional long time scale energy injection from 
their central engines~\cite{dl,du,re,pa}.
In some model, a millisecond pulsar with strong magnetic field can be produced
at birth of a GRB. As the fireball expands, the central pulsar can continuously
supply energy through magnetic dipole radiation. Initially, the energy supply is
rather small, the afterglow shows declining. As it becomes important, the afterglow
shows rising. However, the magnetic dipole radiation should itself attenuate later.
Thus, the down-up-down shape would appear naturally. Dai \& Lu~\cite{dd} further analysed
GRB980519, 990510 and 980326, with dense environment also being taken into
account, and found results agreeing well with observations.

Recently, the GRB000301c afterglow shows three break
appearance in the R-band light curve, and extremely steep decay slope $-3.0$ 
at late time. This unusual afterglow can be explained by assuming more 
complicated additional energy injections and dense medium~\cite{aa}.

Though synchrotron radiation is usually thought to be the main radiation mechanism, 
however, under some circumstances, the inverse Compton scattering may play an 
important role in the emission spectrum, and this may influence the temporal properties 
of GRB afterglows~\cite{we,wl,ii}. Wang, Dai \& Lu~\cite{wg}
even consider the inverse Compton
scattering of the synchrotron photons from relativistic electrons in the
reverse shock. Under appropriate physical parameters of the
GRBs and the interstellar medium, this mechanism can excellently account for
the prompt high energy gamma-rays detected by EGRET, such as from GRB930131.

As some GRBs showed their isotropic radiation energy to be as high as 
$\sim $M$_{\odot}$c$^2$, this has been regarded as an energy crisis.
A natural way to relax this crisis is to assume that the radiation of GRB 
is jet-like, rather than isotropic. However, we should find out its observational 
evidences. Rhoads~\cite{rh,ro} analysed this question, and predicted that the 
sideways expansion in jet-like case will produce a sharp break in the GRB afterglow
light curves (see also Pugliese et al.~\cite{pu}, Sari et al.~\cite{sa} and 
Wei \& Lu~\cite{wu,ww}). Kulkarni et al.~\cite{ku}
regarded the break in the light curve of GRB 990123 as the evidence for jet. 
However, Panaitescu \& M\'{e}sz\'{a}ros~\cite{pn}, Moderski, Sikora \& Bulik~\cite{mo}
performed numerical calculation and denied the appearence of such a sharp break.
Wei and Lu~\cite{wu} re-analysed 
the dynamical evolution of the jet blast wave and found that a sharp break 
can only exist in the case of extremely small beaming angle. Recently, Huang et al.~\cite{uu,nn} 
made a detailed calculation and proved that the breaks in the lightcurves are 
mainly due to the relativistic to non-relativistic transition, not due to edge effect and lateral
expansion effect of the jet, and may appear only
for small electron energy fraction and small magnetic energy fraction. However, they stressed that 
the afterglows of jetted ejecta can be clearly characterized by rapid fading 
in the non-relativistic phase with index $\alpha \ge 2.1$..~\cite{ua}

Gou et al.~\cite{go} used a set of refined dynamical equations and a realistic lateral 
speed of the jet, calculated the evolution of a highly collimated jet that expands in a 
stellar wind environment and the expected afterglow from such a jet. They found that 
in the wind environment, no obvious break will appear even at the time when the 
blast wave transits from the relativistic phase to the non-relativistic phase, and 
there will be no flattening tendency even up to $10^{9}$ s.

\section*{Acknowledgments}
This work is supported by the National Natural 
Science Foundation of China.

\end{document}